\documentclass{article}

\usepackage{arxiv}

\usepackage[utf8]{inputenc} 
\usepackage[T1]{fontenc}    
\usepackage{url}            
\usepackage{amsmath}
\usepackage{booktabs}       
\usepackage{amsfonts}       
\usepackage{nicefrac}       
\usepackage{microtype}      

\usepackage{graphicx}
\usepackage[colorinlistoftodos]{todonotes}
\usepackage[colorlinks=true, allcolors=blue]{hyperref}
\usepackage{verbatim}
\usepackage{url}

\newcommand*{\rood}[1]{#1}
\newcommand*{\blauw}[1]{#1}

\title{Development of a Fragment-Based Machine Learning Algorithm for Designing Hybrid Drugs Optimized for Permeating Gram-Negative Bacteria}

\author{
  Rachael A.~Mansbach \\
  Department of Theoretical Biology and Biophysics\\
  Los Alamos National Lab\\
  Los Alamos, NM \\
   \And
  Inga V.~Leus \\
  Department of Chemistry and Biochemistry\\
  University of Oklahoma\\
  Norman, OK \\
  \And
  Jitender Mehla\\
  Department of Chemistry and Biochemistry\\
  University of Oklahoma\\
  Norman, OK\\
  \And
  Cesar A.~Lopez\\
  Department of Theoretical Biology and Biophysics\\
  Los Alamos National Lab\\
  Los Alamos, NM \\
  \And
  John K.~Walker\\
  School of Medicine\\
  Saint Louis University\\
  St.~Louis, MO\\
  \And
  Valentin V.~Rybenkov\\
  Department of Chemistry and Biochemistry\\
  University of Oklahoma\\
  Norman, OK\\
  \And
  Nicolas W.~Hengartner\\
  Department of Theoretical Biology and Biophysics\\
  Los Alamos National Lab\\
  Los Alamos, NM \\
  \And
  Helen I.~Zgurskaya\\
  Department of Chemistry and Biochemistry\\
  University of Oklahoma\\
  Norman, OK\\
  \And
  S.~Gnanakaran\thanks{To whom correspondence may be addressed}\\
  Department of Theoretical Biology and Biophysics\\
  Los Alamos National Lab\\
  Los Alamos, NM \\
  \texttt{gnana@lanl.gov}
}
\begin{document}
\maketitle

\begin{abstract}
Gram-negative bacteria are a serious health concern due to the strong multidrug resistance that they display, partly due to the presence of a permeability barrier comprising two membranes with active efflux. New approaches are urgently needed to design antibiotics effective against these pathogens.  In this work, we present a novel topological fragment-based approach (``Hunting Fragments Of X'' or ''Hunting FOX'') to rationally ``hunt for'' chemical fragments that promote compound ability to permeate the outer membrane.  Our approach generalizes to other drug design applications. We measure minimum inhibitory concentrations of compounds in two strains of \emph{Pseudomonas\ aeruginosa} with variable permeability barriers and use them as an input to the Hunting FOX algorithm to identify molecular fragments responsible for enhanced outer membrane permeation properties and candidate molecules from an external library that demonstrate good permeation ability.  Overall, we present proof of concept for a novel method that is expected to be valuable for rational design of hybrid drugs.\end{abstract}

\section{Introduction}
In 1950, it was possible to deliver 30 drugs to the market for US \$ 1 billion (using 2016 prices); today not even a single drug can be delivered for that cost \cite{Griffen2018CanIntelligence}.  The number of new molecular entities delivered by the global pharmaceutical industry has remained constant at 16 per year, leading to a severe decrease in productivity or product/cost. One reason for this significant bottleneck in drug design is the increasing slowness with which drugs move through the pipeline from lead identification to clinical trials \cite{Pammolli2011TheRampD,Hay2014ClinicalDrugs}, which is partly a consequence of the dwindling of an ``obvious'' space to search for new drugs.  
Despite the setbacks caused by the increasing attrition rate of potential drug candidates, we nonetheless do possess a large amount of screening data on both approved drugs and failed compounds that are rich in chemical and physical characterization \cite{Kogej2013BigCase}. By leveraging such knowledge, data-driven screening guided by diverse machine learning (ML) methods is emerging as a powerful tool to combat the problem of a dwindling drug pipeline \cite{Lu2017DataOptimization,Reutlinger2014Multi-ObjectivePrioritization,Lavecchia2015Machine-learningApplications} and there is a growing effort to repurpose more data from failed leads to expand what is available \cite{Strittmatter2014OvercomingTricks,King2018SPIDR:Repurposing.}.

In this study, we consider a ML approach to identify chemical fragments and drugs within a molecular space of individual compounds that track activity of interest \cite{Mignani2016CompoundSituation}.  Motivated by the recent use of natural language processing techniques for the analysis of antimicrobial peptides \cite{Cipcigan2018AcceleratingInteractions}, we consider the problem of defining and exploiting a chemical vocabulary--in spirit akin to the ``n-grams'' employed in natural language processing applications \cite{Broder1997SyntacticWeb}--for drug design of small molecules and call our algorithm ``Hunting FOX'' for ``Hunting Fragments Of X.''  From a medicinal chemistry perspective, such an approach potentially provides a pragmatic way to bring together active fragments or diverse sub-molecular spaces in a rational ML-directed manner as building blocks \cite{Goldberg2015DesigningQuality} for novel hybrid drugs. 

We base our approach on the hypothesis that the class of small molecule drugs--both those already in use and those that have been screened in earlier parts of the drug discovery pipeline--contain fragments that confer specific types of activity \cite{Domalaon2018AntibioticPathogens}. We employ our approach for the urgent question of the design and identification of novel compounds with action against Gram-negative bacteria, specifically focusing on identifying fragments and compounds  exhibiting the ability to permeate the outer membrane that can be used as starting points for potential new drug candidates.

 Gram-negative bacteria, such as \emph{P.\ aeruginosa} and \emph{Acinetobacter baumanii}, are notorious for their antibiotic resistance due to the combined effect of their multiple acquired and intrinsic resistance mechanisms. In particular, the sieving mechanism of their double bilayer membranes and the efflux pumps embedded in the inner membrane that actively remove compounds from the cell interior \cite{Arzanlou2017IntrinsicBacteria.,Zgurskaya2015PermeabilityIt,Richter2018TheAntibiotics,Schuster2017ContributionIsolate.,Zgurskaya2018Trans-envelopeBarrier,Cama2019BreachingMembranes} are recognized as the major hurdles in the discovery of new antibiotics and their optimization for clinical use. 
 
 A lack of holistic understanding of the microscopic mechanisms behind antibiotic resistance in Gram-negative bacteria has stimulated development of novel algorithms and studies that (i) identify key components of drugs contributing to bypassing different aspects of multidrug resistance and (ii) allow for rational identification of novel therapeutic leads \cite{Pawlowski2016EvolvingDiscovery,Silver2016AEntry,Richter2018TheAntibiotics}. For example, Wang et al \cite{Wang2016Computer-aidedInfections} used a combination of multiple regression modeling techniques and molecular dynamics simulations to rationally design four antimicrobial peptides (AMPs) with potency against \emph{S.\ aureus}, \emph{P.\ aeruginosa}, and \emph{E.\ coli}, and Lee et al \cite{Lee2018MachinePeptides} used a support-vector-machine-based analysis to not only identify promising AMP candidates from a large online library but also to explain the features that were relevant in endowing them with their antimicrobial properties. A straightforward computational/experimental approach identified that antibiotics designed to accumulate in Gram-negative bacteria should be amphiphilic, rigid, with low globularity and containing an amine \cite{Richter2018TheAntibiotics}, and a random-forest guided approach identified the important physicochemical descriptors of antibiotics for their activities in \emph{E.\ coli} and \emph{P.\ aeruginosa} as electrostatic and surface area related variables versus topological properties, respectively, as a step towards the rational design of drugs with highly specific action against those two pathogens \cite{Cooper2018MolecularAeruginosa}. In addition, numerous high-throughput experimental studies have been performed to identify physicochemical properties of good antibiotics; the consensus has been that small size and lipophilicity are important, although deviations from this have also been identified \cite{Brown2014TrendsPathogens,OShea2008PhysicochemicalDiscovery,Krishnamoorthy2017SynergyBacteria.,Graef2016TheMembrane,Macielag2012ChemicalUniqueness}.
 
On a more experimentally-driven level, the creation of novel hybrids as a route towards more potent antibiotics has recently been gaining traction.  Building on the success of combination drug cocktails, hybrids were initially conceived of as a way to mitigate difficulties arising from the differing pharmacokinetic profiles of two or more drugs that are to be delivered to the bacteria. For example, some promising recent work has shown good antibiotic activity of fluoroquinolone hybrids \cite{Pokrovskaya2009DesignAntibiotics,Wang2014NovelAgent,Xiao2014DesignMicroorganisms}, as well as demonstrating that such hybrids can adopt novel properties not seen in the fragments that make them up: tobramycin-containing hybrids, although not themselves active against Gram-negative bacteria, are emerging as potent adjuvants for other antibiotics \cite{Gorityala2016AdjuvantsEfficacy}.

In this article, we apply Hunting FOX to identify fragments relevant to compound permeation into \emph{P.\ aeruginosa} as a step towards the rational design of novel antibiotic hybrids.  We specifically employ a set of data collected from measuring minimum inhibitory concentrations (MICs) of a $\sim$1,300 compound library enriched with antibacterial compounds and two series of inhibitors of efflux pumps. MICs were analyzed in strategically designed mutant strains of the Gram-negative bacterium \emph{P.\ aeruginosa} with varying properties of outer membrane permeability and efflux efficiency that were previously employed successfully to identify four different classes of antibiotics affected differently by the synergy between the outer membrane and the efflux pumps \cite{Krishnamoorthy2017SynergyBacteria.}.   Based on the ratio of MICs of different compounds in these mutants that are indicative of specific activities as response variables, the Hunting FOX algorithm is used to identify a subset of fragments expected to be relevant and relatively general in predicting permeation behavior. We employ this subset to train classifiers that identify compounds as effective or ineffective permeators.  This simple fragment-based methodology generalizes easily to any response variable of interest: for example, one might classify drugs based on their ability to inhibit the growth of cancerous cells compared to that of normal cells.

\section{Results and Discussion}
\label{sec:results}

In designing the Hunting FOX algorithm and the overall approach described in this article, our objectives were (i) to develop a vocabulary of submolecular fragments, similar to n-grams in natural language processing \cite{Broder1997SyntacticWeb}, to predict drug efficacy in permeating the outer membrane, and (ii) to validate our approach on a large external drug database with potential drug discovery implications.  In the following section, we outline the algorithm and demonstrate the identification of promising submolecular fragments for use in antibiotic hybrid design. We show a rationalization for why such fragments would be identified.  Finally, we experimentally validate that  classifiers trained on the relevant fragments possess significant ability to identify hits from an external library of drug compounds not contained in the original data set. 

\subsection{Overview of the Hunting FOX Algorithm}
\label{subsec:foxoverview}
\begin{figure}
    \centering
    \includegraphics[width=\textwidth]{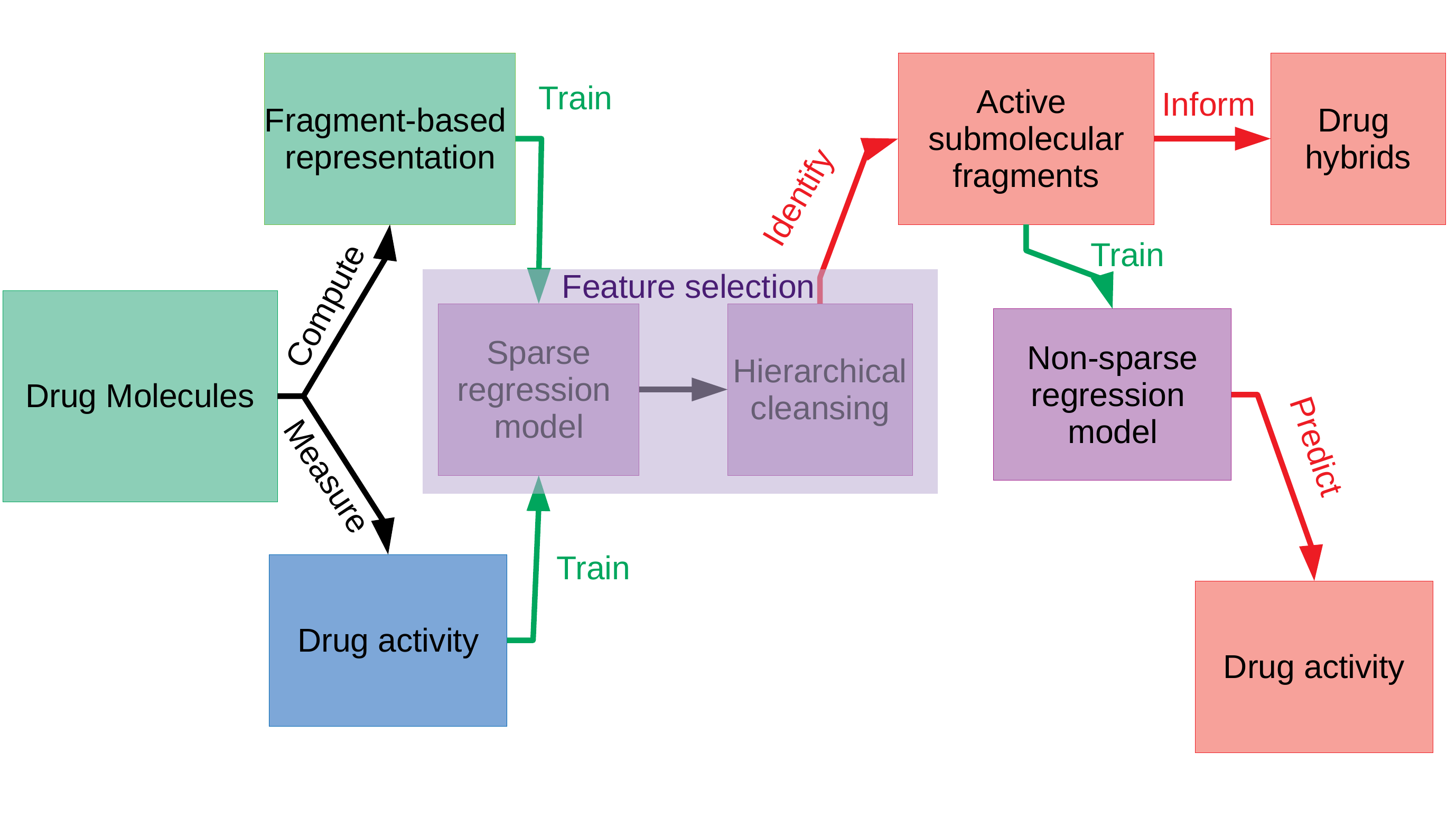}
    \caption{Schematic of the basic Hunting FOX algorithm. We compute a fragment-based representation of drugs and use a combination of sparse regression and a hierarchical cleansing proceed to select a subset of relevant fragments, which are expected to be relevant to the design of novel antibiotic hybrids. We use these fragments to train a non-sparse regression model, from which we may predict drug class for novel molecules.  In this study, the drug activities (blue) employed were MIC ratios of compounds in two different mutant strains of \emph{P.\ aeruginosa} PAO1. The algorithm used these MIC ratios to classify a set of compounds based on their ability to permeate the outer membrane.}
    \label{fig:foxschematic}
\end{figure}

\begin{figure}[h]
    \centering
    \includegraphics[width=\textwidth]{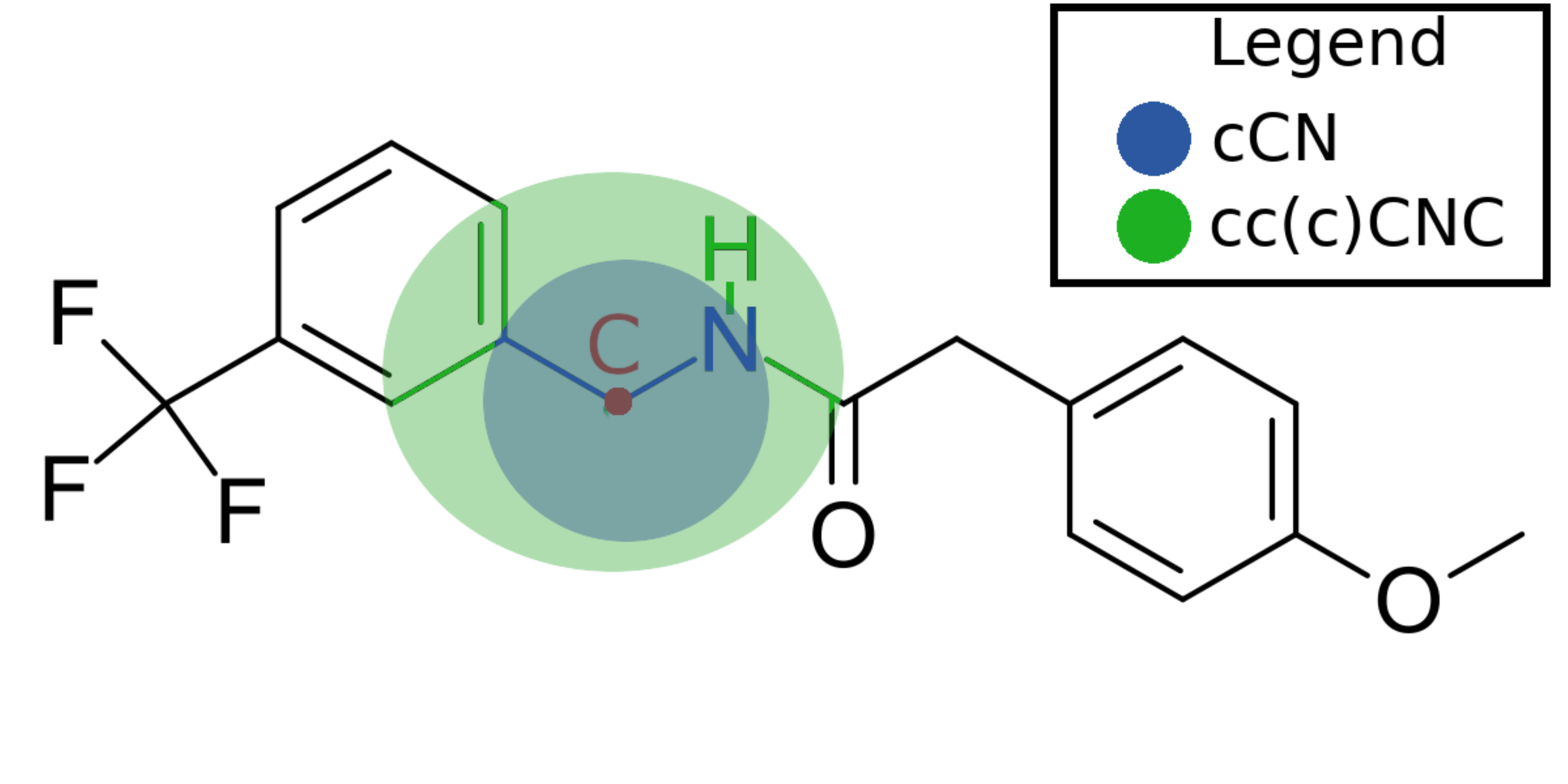}
    \caption{Example of fragment definition by radius. The figure shows an arbitrarily chosen compound (OU-31237) with an arbitrarily chosen central atom in red. We show both the fragment of radius 1 bond about the central atom (in blue) and the fragment of radius 2 bonds about the central atom (in green, also includes the blue bonds and atoms). In the legend we indicate the SMILES string associated with each fragment. In \blauw{Table S1} we list all fragments of radius 1 and 2 contained in this molecule. Visualization of molecule chemical structure was rendered employing tools from the CDD Vault from Collaborative Drug Discovery (Burlingame, CA. \url{www.collaborativedrug.com}) \cite{Hohman2009NovelDiscovery}}
    \label{fig:radexample}
\end{figure}

The overall workflow of the Hunting FOX algorithm is shown in \blauw{Fig.\ \ref{fig:foxschematic}}. We begin by defining a fragment-based representation for the molecules.  For each molecule in the dataset, we identify all fragments that comprise it (see \blauw{Fig.\ \ref{fig:radexample}} and \blauw{Sec.\ \ref{subsubsec:rep}} for details). We gather all fragments in all molecules for a total of $N_f$ fragments and represent each molecule $M$ as a $N_f$-length vector of frequencies, where every entry is the number of times a particular fragment appears (may be 0), divided by the number of atoms in molecule $M$, $L(M)$. Although this is a somewhat naive representation, we employ it as an initial way to test whether information about permeation ability is contained within the fragments that compose a set of compounds.

Next, we perform feature selection by using sparse regression and $k$-fold cross-validation on the experimental data of our choice \cite{Kohavi1995ASelection}. In other words, we divide our experimental data into $k$ fractions and hold out 1/$k$ for testing of each model, training it on the other $k-1$/$k$ data. The models may be linear--in the case of continuous data--or logistic (that is a classifier)--in the case of discrete data. By using the LASSO regularization technique \cite{Tibshirani1996RegressionLasso}, which zeros out variables in the original representation that are not strongly correlated with the outcomes, we are able to identify a subset of the fragments that are predictive of the experimentally-measured drug activity. We then perform a post-processing step to eliminate hierarchical relationships in the retained set of molecules, which results in a set of active submolecular fragments, none of which are submolecules or supermolecules of any of the other identified fragments in the set.  

Having identified the most important and non-redundant set of fragments that are predictive of the desired experimental outcome, we may use these fragments to inform the design of novel drug hybrids.  In addition, we train a second set of $k$ regression models, either linear or logistic, on the experimental drug activity data, and use them as predictors of the activity of novel compounds on which they were not trained.  Currently, we expect to find best results by running the algorithm repeatedly on different $k$-fold splits and allowing the models to vote, thus leveraging the power of ensemble learning \cite{Dietterich2000EnsembleLearning}.

\subsection{Input to the Hunting FOX algorithm: MIC ratios as permeation indicators}
\label{subsec:studydesign}
Two primary mechanisms of multidrug resistance in Gram-negative bacteria are the highly impermeable outer membranes, which prevent drug incursion into the cell, and the efflux pumps, which actively extrude those drugs that do successfully navigate the outer membrane. To separate these effects, we have recently developed techniques to create different mutant strains of Gram-negative bacteria \cite{Krishnamoorthy2016BreakingMembrane.,Krishnamoorthy2017SynergyBacteria.}. In this study we primarily used two mutants of \emph{P.\ aeruginosa} PAO1 strain: the ``P$\Delta$6'' mutant, and the ``P$\Delta$6-Pore'' mutant \cite{Cooper2018MolecularAeruginosa}.  The ``P$\Delta$6'' mutant is a variant of \emph{P.\ aeruginosa} in which the genes encoding for the six best characterized efflux pumps have been deleted, which essentially removes the contribution of active efflux in antibacterial activities of antibiotics. The ``Pore'' mutant is a variant in which mutation of the outer membrane makes it hyperporinated;  that is, modified to express large, nonspecific pores that allow nondiscriminate entry of drugs, which essentially removes the effects of the impermeable outer membrane.  We have previously demonstrated the use of this technique in \emph{E.\ coli} and \textit{P. aeruginosa} \cite{Krishnamoorthy2016BreakingMembrane.,Krishnamoorthy2017SynergyBacteria.,Cooper2018MolecularAeruginosa}. The ``P$\Delta$6-Pore'' mutant is a variant combining both previous mutations.  For the drug property input to the algorithm (see \blauw{Fig.\ \ref{fig:foxschematic}}), we experimentally measured the MICs of over 1,300 compounds, out of which more than five hundred drugs exhibited antibacterial activities at least in one out of two different mutant strains of \emph{P.\ aeruginosa} PAO1. We then compute the ratio of compound MIC values in the P$\Delta$6-Pore mutant of \emph{P.\ aeruginosa} PAO1 to their MIC values in the P$\Delta$6 mutant of \emph{P.\ aeruginosa} PAO1.  When this ratio goes to one, there is no effect of the outer membrane on the efficacy of the drug in question, so we say that the drug permeates well.  We employ this experimental data as input to the Hunting FOX algorithm as described in the previous section.

\subsection{Hunting FOX discovers active submolecular fragments for hybrid drug design}
\label{subsec:topfrags}

\begin{figure}
    \centering
    \includegraphics[width=\textwidth]{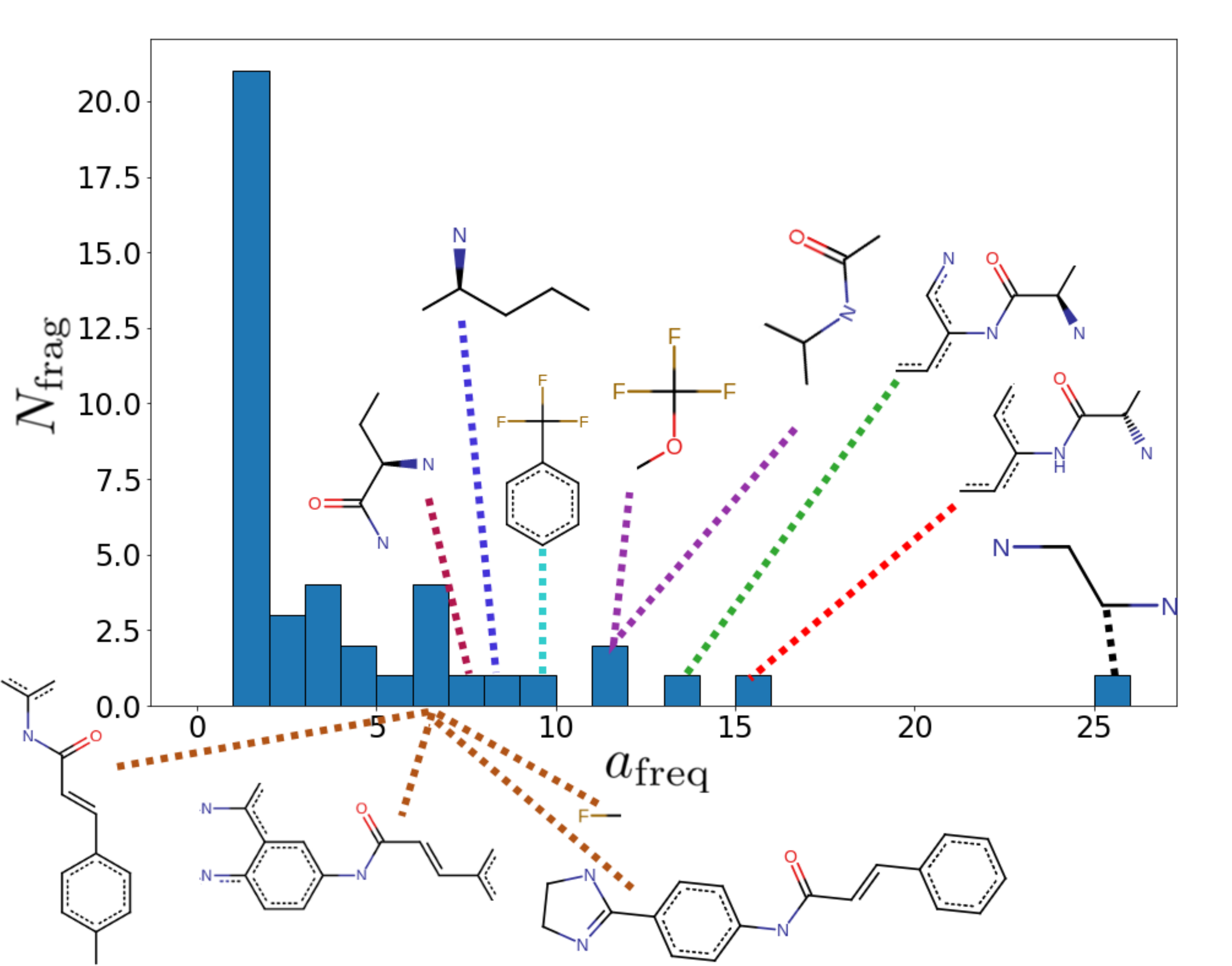}
    \caption{Histogram showing the number of fragments $N_\textrm{frag}$ identified by $a_\textrm{freq}$/\rood{28} runs of Hunting FOX with randomly shuffled training/testing data. We render the \rood{12} fragments that are reported by more than five runs and identify which bin they fall into. The fragments were rendered using Marvin 19.16.0, 2019, ChemAxon (\url{http://www.chemaxon.com}). We do not show hydrogen occupancy as it may change depending on how a fragment is connected to a molecule. Different colored dotted lines indicate fragments lying in different bins.}
    \label{fig:uncorrweb}
\end{figure}

We run the Hunting FOX algorithm \rood{twenty-eight} times starting from a different random seed each time, which corresponds to a different split for the $k$-fold training data, and each time identify a set of important molecular fragments. In \blauw{Fig.\ \ref{fig:uncorrweb}}, we report the number of fragments out of 43 fragments that appear between one and \rood{twenty-eight} times, while in \rood{Supporting File \texttt{anc/topfragments.xlsx}} we report all identified fragments and their frequency of appearance. We note a moderate level of robustness in the algorithm: although no fragments are reported by every run, \rood{one out of forty-four} does appear in \rood{twenty-six of twenty-eight} runs, and \rood{twenty-two of forty-three} fragments appear in more than one run.  The lack of more fragments appearing repeatedly indicates a dependency on the training data that can likely be ameliorated by employing a more robust and less naive representation (cf.\ \blauw{Sec.\ \ref{subsec:limitations}}); however, even moderately performing models experience a significant performance enhancement when combined \cite{Dietterich2000EnsembleLearning}.

Furthermore, although it is beyond the scope of this paper to employ these fragments identified for experimental hybrid synthesis, there are several notable chemical features of the fragments that provide \emph{a posteriori} empirical support for our procedure.  First, we note the appearance of a trifluoromethyl fragment (\blauw{Fig.\ \ref{fig:uncorrweb}}, central molecule at $a_\textrm{freq}=9$) in nine out of twenty-eight runs. This particular functional group is currently widely used, not only due to its electronic properties that improve synthesizability, but also in terms of biomolecular inhibition \cite{Kelly2013TrifluoromethylApplication}. In fact, a number of recent promising antibiotic hybrids have contained trifluoromethyl fragments, aiming to target bacterial proteases \cite{Wang2014Fluorine20012011}. Second, twenty-four of forty-four of all fragments and six of twelve fragments appearing more than \rood{five times} contain part of a benzene ring or one or more whole benzene rings. Such fragments have recently been demonstrated to improve permeability \cite{Richter2017PredictiveAntibiotic}, and this result also highlights that the predictive algorithm is able to discriminate 3-dimensional features that are not directly employed in our code. In fact, the presence of rigid benzene rings dramatically improves both rigidity and globularity, important for membrane translocation \cite{Richter2017PredictiveAntibiotic}. Third, a large number of the fragments identified contain primary and secondary amine groups. Such groups are expected to strongly interact with both the KDO (2-Keto-3-deoxy-octonate) and lipid-A regions of the LPS, functioning as specific anchors for highly anionic membranes (e.g. bacterial outer membrane) \cite{Savage2002AntibacterialAntibiotics}. Indeed, it is well-known that including such groups does improve intracellular accumulation \cite{Richter2017PredictiveAntibiotic} and specificity for Gram-negative bacteria \cite{Moretti2019CationicMembranes}. It is remarkable that we were able to capture such behavior without leveraging chemical expertise in the initial steps.

\subsection{Hunting FOX identifies antibiotic candidates from an external library of molecules}
\label{subsec:exphits}

We now demonstrate that our algorithm produces models sufficient to identify novel molecules with desired properties from a library of molecules with unknown properties using only the fragments identified in the previous section.  Employing a library of 30,929 molecules that were not part of the training or testing set for any of the regression models, we used the non-sparse regression models of 28 repeated randomized iterations of the Hunting FOX algorithm to identify which of these molecules might be expected to display outer membrane permeation properties as measured by the MIC ratios $\frac{\mu_{P\Delta 6\textrm{-Pore}}}{\mu_{P\Delta 6}}$. In each iteration, we reported as potential ``hits'' those molecules that were predicted by all $k$ non-sparse models to have MIC ratios of $\frac{\mu_{P\Delta 6\textrm{-Pore}}}{\mu_{P\Delta 6}}>0.8$.  .  

\begin{figure}
    \centering
    \includegraphics[width=\textwidth]{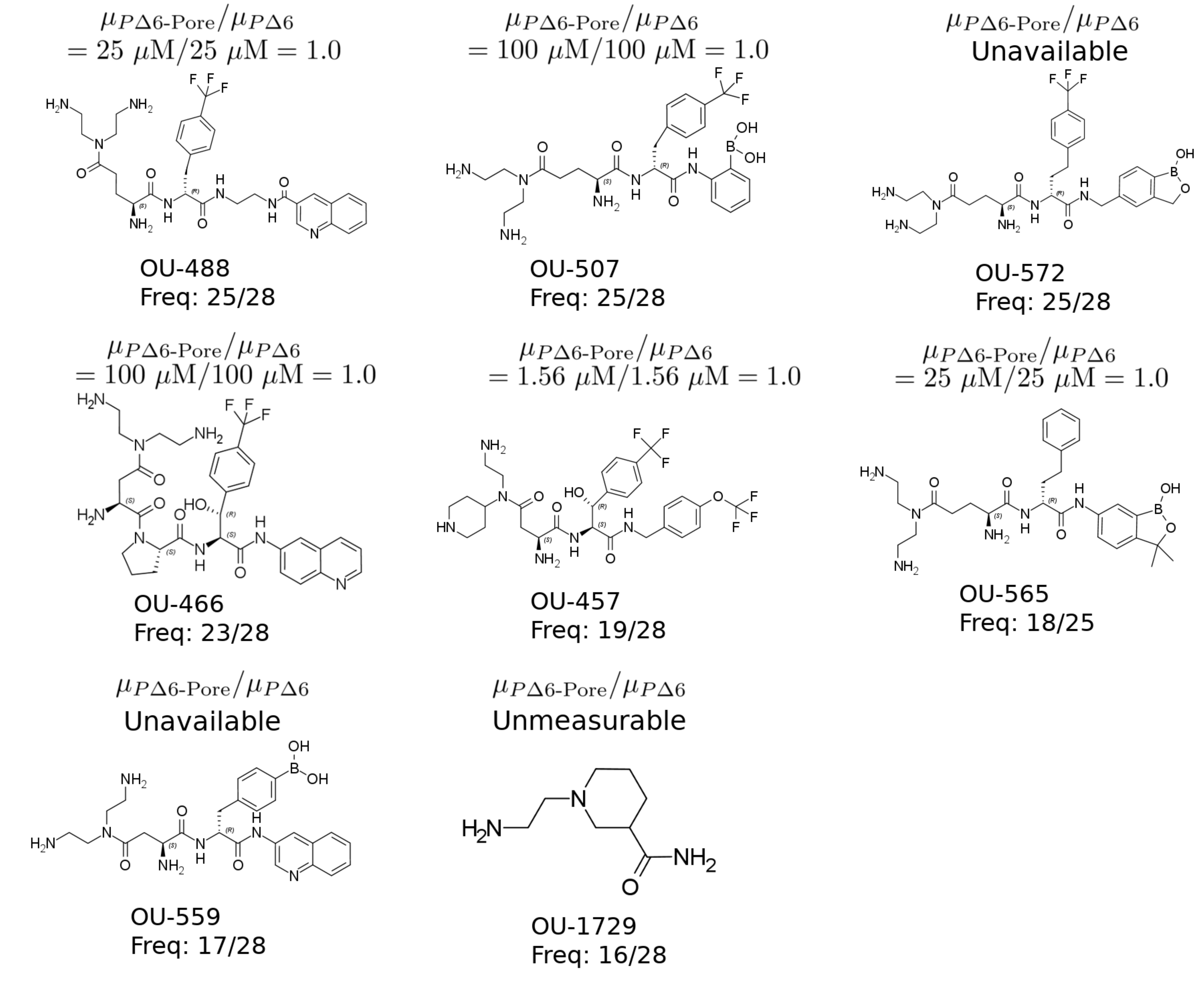}
    \caption{Molecules selected as hits by more than \rood{14/28} repeated runs of the Hunting Fox algorithm. We note the experimental measurements above the chemical structures of the molecules and the molecule name below.  Visualizations of molecule chemical structure were rendered employing tools from the CDD Vault from Collaborative Drug Discovery (Burlingame, CA. \url{www.collaborativedrug.com}) \cite{Hohman2009NovelDiscovery}.}
    \label{fig:tophits}
\end{figure}

Eight compounds were identified by at least 50\% of the repeated runs of the Hunting FOX algorithm (\blauw{Fig.\ \ref{fig:tophits}}).  Among these top eight identified compounds, \rood{five} possessed antibacterial activities and their permeation properties could be assessed by measuring MICs in P$\Delta$6 and P$\Delta$6-Pore strains . For all these compounds, the ratio of MICs $\frac{\mu_{P\Delta 6\textrm{-Pore}}}{\mu_{P\Delta 6}} \approx 1.0$ indicating that they are good permeators.  Compounds OU-572 and OU-559 were unavailable, but belong to the same structural series as OU-457 and OU-466, sharing with them certain structural fragments, and are likely to have similar properties.  The remaining compound OU-1729 is not expected to display measurable MICs and their permeation could not be assessed using growth inhibition assays. The difficulty of measuring outer membrane permeation in isolation limits our ability to estimate the overall performance of the model on the external library. However, the fact that all drugs with measurable antibacterial activities were hits provides experimental validation of our models and, more broadly, of our algorithm.


Of the \rood{eight} top predicted hits, \rood{five} contain the (trifluoromethyl)benzene functional group mentioned previously (cf.\ \blauw{Sec.\ \ref{subsec:topfrags}}). Nowadays, fluorine containing compounds are synthesized in pharmaceutical research on a routine basis and about 10 percent of all marketed drugs contain a fluorine atom. The major rationale is that the presence of fluorine atoms in biologically active molecules can enhance their lipophilicity and thus their uptake and transport. In particular, the trifluoromethyl group (-CF3) confers increased stability and lipophilicity in addition to its high electronegativity. This enrichment with (trifluoromethyl)benzene-containing compounds is somewhat to be expected, as our initial dataset contained a not-insubstantial number of compounds with this functional group. This provides an additional check of our algorithmic approach, but also points to a potential limitation: it is unable to assess fragments that do not appear in the training set, which limits its ability to generalize to wholly novel molecules.

Overall, we demonstrate that the Hunting FOX algorithm is capable of achieving the goals laid out for it: it provides an approach that allows the reuse of components of compounds that failed initial screening for particular applications, and, in the process of doing so, it specifically (i) identifies a set of fragments containing information about compound ability to permeate the outer membrane of \emph{P.\ aeruginosa}, which are of great interest for the design and synthesis of novel antibiotic hybrids with action against Gram-negative bacteria, and (ii) identifies a set of compounds from an external library expected to be good membrane permeators, providing experimental validation of the informational content of the fragments.

\subsection{Current limitations and future improvements}
\label{subsec:limitations}
One major difficulty with the Hunting FOX algorithm stems from the fact that fragments are combinatorial in nature, which means that to describe a larger chemical library a huge number of fragments will be necessary. Due to the sparsity of the representation, it is difficult to apply standard statistical techniques to the regression results. Furthermore, we note that the fragments identified and hence the molecules that the final classifiers identify change not insignificantly when the training/testing split is changed.  Of \rood{993} molecules were identified as hits by one or more runs of the Hunting FOX algorithm as potential outer membrane permeators (see Supplementary File \texttt{anc/projected\_hits.xlsx} and \blauw{Fig.\ S1}) only 326 (33\%) were reported in more than one run of the algorithm and only eight (0.8\%) were reported in more than half of the runs. This observation implies that the current representation is insufficiently robust to changes in the dataset, which we hypothesize is a consequence of (i) magnification of noise due to an overly high-dimensional representation, (ii) lack of incorporation of hierarchical and locational relationships among fragments in the initial construction of the representation, and (iii) uncertainties experimental measurements of MICs. This is similar to the difficulties encountered when attempting to employ one-hot-encoding to learning on textual data.

With respect to fragment representation, work on word and sentence embeddings provide a good source of how to compress highly sparse, combinatorial representations in a meaningful way; combining, for example, the word2vec \cite{Mikolov2013DistributedCompositionality,Mikolov2013EfficientSpace,Mikolov2013LinguisticRepresentations} or GloVE \cite{Pennington2014GloVe:Representation} approach with an autoencoder approach \cite{Gomez-Bombarelli2018AutomaticMolecules} would provide a compact representation that is reconstructible with the fragment composition, as well as potentially identifying a latent space of meaningful variables.  Similar approaches have begun to show promising results in the general area of \emph{de novo} drug design \cite{Ozturk2018APrediction,Schwaller2018ChemicalModels}. An additional limitation of the current representation is that it contains geometric information only implicitly, which leads to difficulty in the direct assessment of relevant fragment properties and the potential of losing important information such as enantiomerism. Therefore we also plan to pursue the incorporation of three-dimensional information into our representation, following early studies in \emph{de novo} drug design in that area \cite{Elton2019DeepArt}.

Additionally, as the discussion in \blauw{Sec.\ \ref{subsec:exphits}} highlighted, our experimental assays are currently insufficient to identify molecules that permeate well but do not have antibiotic action, a necessary development to test the full suite of predicted hits from the Hunting FOX algorithm, as well as a necessary step in the understanding of mechanisms of antibiotic resistance in isolation.  Development of direct assays to analyze permeation across the outer membrane of \emph{P.\ aeruginosa} is in progress.

\section{Methods}
\label{sec:methods}


In this section, we describe the details of the Hunting FOX algorithm as applied in this article (see also \blauw{Sec.\ \ref{subsec:foxoverview}} for an overview). First, we detail the fragment-based molecule representation, followed by the details of MIC measurement and ratio computation (see also \blauw{Sec.\ \ref{subsec:studydesign}}).  Next, we describe the feature selection step, in which for this application we employ sparse multinomial logistic regression, followed by hierarchical cleansing on an interpretable subset of the fragments that were retained after sparse classification.  Finally, we describe the training and performance of a second set of (non-sparse) classifiers.

\subsection{Molecular Descriptor Vectors}
\label{subsubsec:rep}
As mentioned in \blauw{Sec.\ \ref{subsec:foxoverview}}, we consider each molecule to be a vector comprised of some number of submolecules. We begin with the two-dimensional representation of a molecule as a set of atoms and bonds connecting the atoms. For every atom in the molecule, we identify all fragments consisting of that central atom plus the atoms that lie within $k$ bonds of it (cf.\ \blauw{Fig.\ \ref{fig:radexample}}).  We consider increasing radii of $k$ bonds from 1 to 10 and identify all unique submolecular fragments within the molecule by using the \texttt{FindAtomEnvironmentOfRadiusN} function in the \texttt{rdkit} Python package \cite{RDKit:Cheminformatics}. We compute a total of $N_f = 22,139$ fragments $x_i$ in all molecules in the set of molecules considered. Due to end effects, some fragments may be identified at more than one radius or central atom; however, we do not consider fragments to be different because they appear at different bond radii--they are defined solely by the atom types, bond types, and topology making them up. Once all such fragments are computed, we represent each molecule $M$ by a vector of fragment frequencies $\vec{\ell}(M) = [f(x_1,M),...,f(x_N,M)]$, where,

\begin{equation}
    f(x_i,M) = n(x_i,M)/L(M),
\end{equation}
in which $n(x_i,M)$ represents the number of times fragment $x_i$ appears in molecule $M$, and $L(M)$ is the size of $M$ computed as the total number of atoms.  Intuitively, containing larger numbers of relevant fragments should be correlated with increased activity; however, we wanted a metric that was independent of molecular size. We included any fragment appearing in any molecule in the set of considered molecules to account for fragment absence as well as presence.

\subsection{Measurement of minimum inhibitory concentrations and data cleaning}
\label{subsubsec:MICs}
\textit{P. aeruginosa} cells were grown in Luria Bertani Broth (LB) (10 g tryptone, 5 g yeast extract, 5 g NaCl per liter, pH 7.0) at 37$^{\circ}$ C with shaking. Minimum inhibitory concentration (MIC) determination was carried out using the 2-fold broth dilution method as described previously \cite{Krishnamoorthy2017SynergyBacteria.}. The expression of the Pore was induced at OD600 ~0.3-0.4 by addition of 0.1 mM IPTG.

In practice, due to experimental limitations, the data must be cleaned. We remove any molecules from our analyses for which both MIC values were too high to measure, as it is impossible to know whether these molecules actually have ratios close to one or not, which results in a dataset of 595 molecules from which we compute the fragments for the descriptor vectors. We also set any ratios wherein the denominator was too high to measure to zero, assuming that such molecules will not contain desirable fragments. Due to the 2-fold error in the MIC measurements, it was possible to compute ratios greater than one, which were set to one to avoid inclusion of erroneous information. Finally, certain measurements were given as ranges of MIC values and, in each of those cases, we chose the mean of the range.

\subsection{Feature selection}

The feature selection portion of the Hunting FOX algorithm consists of two steps. First, we train a set of sparse classifiers that return zero coefficients for unimportant descriptors--in this case, molecular fragments that are not predictive of compound permeability. Second, we perform a postprocessing step in which those fragments that retain non-zero coefficients after sparsification are subsampled further to eliminate hierarchical relationships among them. 

\subsubsection{Multinomial logistic regression for fragment selection and external molecule classification}
\label{subsubsec:classify}
Due to the fact that MIC values are reported in powers of two, the data are quite discrete, and for this reason, we chose to train a classifier with five classes (cf.\ \blauw{Fig.\ S2}).   We define the classes as follows, based on the approximate ``streaking'' in the data, which we expect to indicate drastically different molecular classes.  If $\frac{\mu_{P\Delta 6\textrm{-Pore}}}{\mu_{P \Delta 6}}<0.2$, $\textrm{Class}\left(\frac{\mu_{P\Delta 6\textrm{-Pore}}}{\mu_{P \Delta 6}}\right) = 0$; if $ 0.2\leq \frac{\mu_{P\Delta 6\textrm{-Pore}}}{\mu_{P \Delta 6}} < 0.4 $, $\textrm{Class}\left(\frac{\mu_{P\Delta 6\textrm{-Pore}}}{\mu_{P \Delta 6}}\right)  = 1$; if $0.4 \leq \frac{\mu_{P\Delta 6\textrm{-Pore}}}{\mu_{P \Delta 6}} < 0.6$, $\textrm{Class}\left(\frac{\mu_{P\Delta 6\textrm{-Pore}}}{\mu_{P \Delta 6}}\right)  = 2$; if $0.6 \leq \frac{\mu_{P\Delta 6\textrm{-Pore}}}{\mu_{P \Delta 6}} < 0.8$, $\textrm{Class}\left(\frac{\mu_{P\Delta 6\textrm{-Pore}}}{\mu_{P \Delta 6}}\right)  = 3$; and if $\frac{\mu_{P\Delta 6\textrm{-Pore}}}{\mu_{P \Delta 6}} > 0.8$, $\textrm{Class}\left(\frac{\mu_{P\Delta 6\textrm{-Pore}}}{\mu_{P \Delta 6}}\right)  = 4$.  The class breakdown is as follows: $\approx$ 48\% of MIC ratios fall into class 0, 10\% into class 1, 9\% into class 2, 10\% into class 4, and 22\% into class 4, where $\mu_{P\Delta 6\textrm{-Pore}}$ is the MIC of the compound in the $P\Delta 6$-Pore mutant and $\mu_{P\Delta 6}$ is the MIC of the compound in the $P\Delta 6$ mutant.

We assume that a large number of the 22,139 fragments comprising the initial set of 595 molecules will not be relevant to the output variables and therefore we initially perform feature selection by seeking a set of sparse classifiers that will retain only a small number of features. A ``sparse classifier'' is a classifier with an additional regularization parameter that can be tuned to find solutions with only a small number of non-zero parameters.  In addition to cutting down on the number of relevant variables, such an approach improves interpretability and helps prevent overfitting, a problem in which a model shows high performance on the dataset for which it is trained, but poor performance when asked to generalize to new data points \cite{Lever2016ModelOverfitting}. 

To fit initial sparse classifiers for a single run of the Hunting FOX algorithm, we employed stratified 5-fold cross-validation followed by multinomial logistic regression (see \blauw{Sec.\ S1}) on 595 molecules out of a library of 31,622 for which $\left(\frac{\mu_{P\Delta 6\textrm{-Pore}}}{\mu_{P \Delta 6}}\right)$ was measured and identified as a definite value as described in \blauw{Sec.\ \ref{subsec:studydesign}}.  The data was curated and organized by employing the CDD Vault from Collaborative Drug Discovery (Burlingame, CA. \url{www.collaborativedrug.com}) \cite{Hohman2009NovelDiscovery}. We employed the \texttt{StratifiedKFold} class with a random seed from the python \texttt{scikit-learn} package \cite{Pedregosa2011Scikit-learn:Python} to split each of the two groups into five training and testing sets wherein the testing sets were disjoint and both training and testing sets approximately reproduced the distribution of different classes of the original group. For each split, we employed the \texttt{cvglmnet} function from \texttt{glmnet-python} package with \texttt{family=`multinomial'}, \texttt{mtype=`grouped'}, and $\alpha=1$ to fit a sparse multinomial logistic regression model on the training set with regularization parameter $\lambda$ controlling the strength of the LASSO penalty for sparsification  \cite{Balakumar2016GlmnetPython}. Employing the built-in cross-validation from the package, we identified $\lambda = \lambda_{min}$ and $\lambda = \lambda_{1se}$ wherein $\lambda_{min}$ corresponds to the regularization parameter with minimum deviance and $\lambda_{1se}$ corresponds to the maximum regularization parameter with deviance within one standard deviation of the minimum. The potential benefit of employing $\lambda_{1se}$ is that it forces the model to be sparser and may therefore reduce overfitting \cite{Balakumar2016GlmnetPython,Friedman2010RegularizationDescent.,Tibshirani2012StrongProblems,Simon2013ARegression}, and all trained classifiers we discuss from now on are those employing $\lambda = \lambda_{1se}$.

\subsubsection{Procedure for cleansing hierarchical relationships}
\label{subsubsec:hiercleanse}
From the set of fragments that result in nonzero coefficients in the model, we retain only fragments that have positive coefficients for prediction of class 4 occupancy--that is, fragments whose existence imply a compound will have a high permeating ability--or negative coefficients for prediction of class 0 occupancy--that is, fragments whose existence imply a compound will have a non-zero permeating ability.  We do not retain fragments with coefficients solely pertaining to the middle three classes for two reasons: (i) the interpretation of such fragments is less clear and (ii) due to the small number of molecules in these classes, the performance of classifiers for prediction of those three classes was significantly worse. This step would not necessarily have to be performed for all applications.  We further sparsify the retained fragments by finding a subset that are not hierarchically related. 

Our five trained sparse classifiers contain two natural ways to assess relative importance of the fragments that form the descriptors of molecules: ``model voting''--that is, in how many of the different classifiers does a fragment appear--and the coefficients that relate a particular fragment frequency to the probability of a molecule occupying a particular class. We expect that a fragment that shows up in more models contains information that is more generally applicable for prediction, while a fragment that appears with a large magnitude coefficient for a particular class represents a fragment that is of greater significance to a particular classifier for prediction of occupancy in that class.  There is, however, an important caveat: due to the fact that we considered a set of variable radii when determining which fragments to include, certain fragments are hierarchically related to one another--that is, they are sub- or super-molecular fragments of one another. In other words, two classifiers may find very similar fragments that are hierarchically related but not identical, which complicates identification of the most important fragments.  Therefore, instead of simply seeking fragments that appear in multiple classifiers with large magnitude coefficients, we instead performed a hierarchical cleansing procedure to produce a minimal set of non-hierarchically related fragments with  largest coefficients that contain information that appears in multiple classifiers. 

Our procedure is as follows: first, consider only fragments with non-zero coefficients in at least one classifier.  Then, find the subset $\{\mathcal{H}\}$ that are hierarchically related within all classifiers, meaning that for any given fragment $x_i \in \{\mathcal{H}\}$ appearing in one classifier, either that same fragment or a fragment that it contains or is contained by appears in every other classifier.  Next, consider fragments in order of increasing coefficient magnitude, where we take the maximum coefficient magnitude of all the coefficients in all classifiers in which the fragment appears. Each time such a fragment is part of a hierarchical relationship with another fragment, remove it from consideration. Continue doing so until all that remains are fragments that are not hierarchically related to one another. This preferentially retains fragments that are ranked as important to at least one classifier and also ensures that all retained fragments contain information present in all classifiers.

\subsection{Classifiers for whole molecule identification}

Once we identified a set of likely active fragments, we trained a new set of non-sparse classifiers using only the non-hierarchically-related subset to validate the informational content contained in the retained fragments by identification of molecules from an external library (cf.\ \blauw{Sec.\ \ref{subsec:exphits}}) and assessment on a test set.

To fit this new set of classifiers, we once again employed stratified 5-fold cross-validation in conjunction with multinomial logistic regression, but with a ridge penalty rather than a LASSO one.  The only difference in the fitting procedure besides using different descriptors was that we employed $\alpha=0$ instead of $\alpha=1$ in the \texttt{cvglmnet} function to change from LASSO to ridge regression. 

\subsubsection{Classifier performance}

 We assessed the performance of the non-sparse classifiers through the twin metrics of receiver operating characteristic (ROC) curves and enrichment on the accompanying test set. The ROC curve is an illustration of the true positive rate, $TPR$, versus the false positive rate, $FPR$, as a function of the probability returned by the classifier that a particular molecule falls into the given class. For a classifier that contains no information, that is whose behavior is the same as random, the plot should lie along the line of equivalence, $TPR = FPR$. For a perfect classifier, the $TPR$ would immediately rise to one \cite{Brown2006ReceiverTutorial}.  We define ``enrichment,'' $\mathcal{E}$, of a class at a level $m$ as,
 \begin{equation}
     \mathcal{E}(m) \equiv \frac{\frac{1}{m}\sum_{j=1}^m y_j^* - \frac{1}{N} \sum_{j=1}^N y_j^*}{\frac{1}{N}\sum_{j=1}^N y_j^*}
 \end{equation}
where $y_j^* \in [0,1]$ is the list of true occupancies for the samples in  a category reordered in the order of decreasing probability as predicted by the classifier, $N$ is the total number of samples, and $\frac{1}{N}\sum_{j=1}^N y_j^*$ is the total fraction of samples that belong to the class. Thus the enrichment at a level $m$ measures the difference between the percentage of hits identified by the classifier from the percentage of hits that would be found by random chance.
 
\begin{figure}
    \centering
    \includegraphics[width=\textwidth]{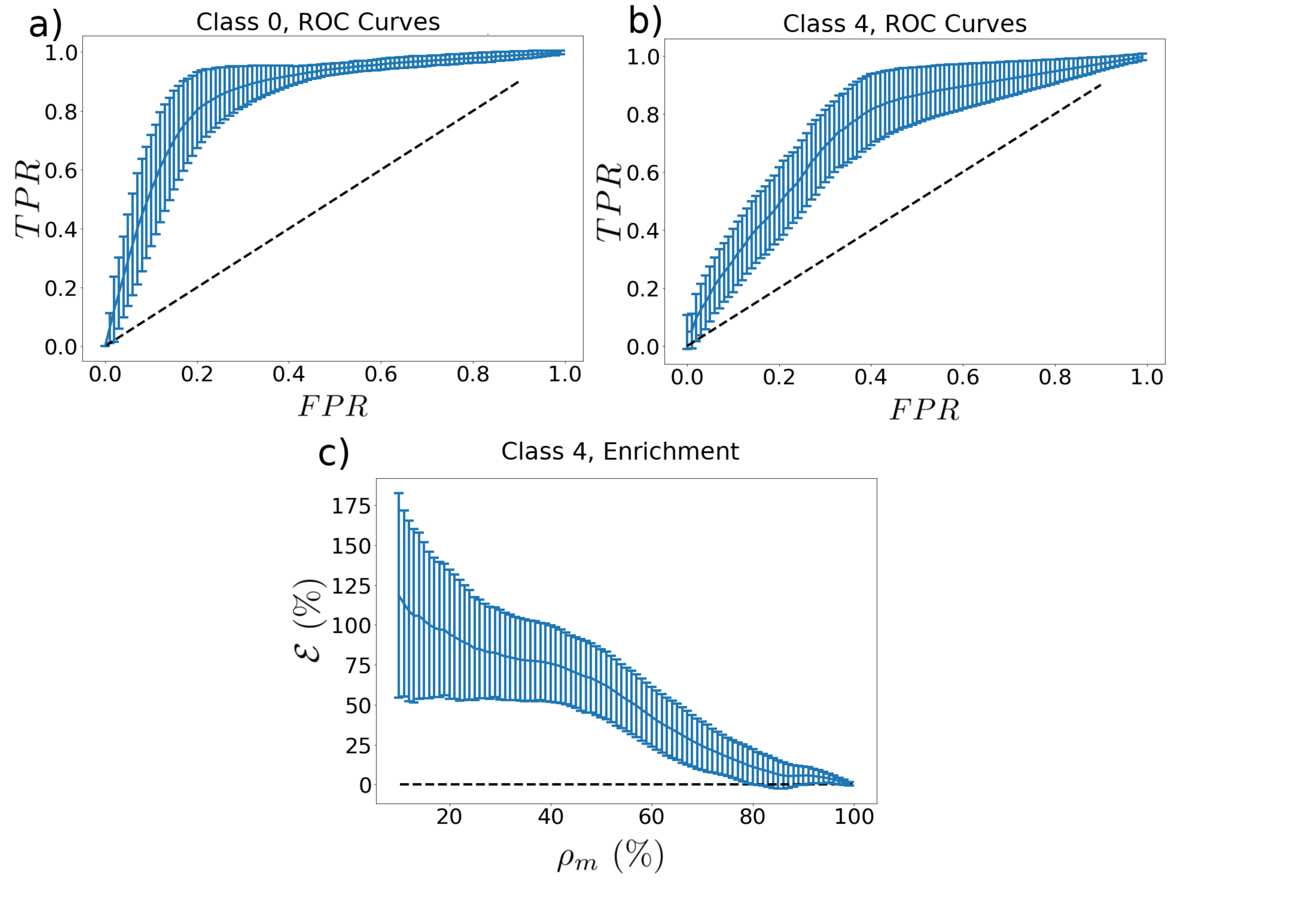}
    \caption{Average ROC curves resulting from different classifiers for (a) class 0 and (b) class 4 and (c) average enrichment ($\mathcal{E}$) versus percent $\rho_m$ of molecules considered after sorting by predicted probability of class 4 occupancy. Curves were linearly interpolated at 100 evenly-spaced intervals and then averaged at those points. Error bars represent standard deviations over 140 classifiers produced from 28 different iterations of the Hunting FOX algorithm to show the range of the data. Dashed black lines indicate the expected performance of a random classifier.}
    \label{fig:avgclassperform}
\end{figure}

 In \blauw{Fig.\ \ref{fig:avgclassperform}}, we show the average performance of \rood{140} classifiers trained in \rood{28} iterations of Hunting FOX starting from different randomized seeds. We linearly interpolate the separate curves to assess performance at the same points and report the average and standard deviation of all 140 classifiers for ROC curves on classes 0 and 4, and enrichment on class 4. We consider only the enrichment of class 4, because that indicates the ability of a classifier to identify highly-permeating molecules, which is of greatest interest in this work.   In \blauw{Fig.\ S3}, we also show the performance of all 140 classifiers separately.  The performance on classes 1-3 (not shown) is no better than random, as expected, because in the previous step we retained only fragments related to class 0 or class 4. Although we could have trained a 2-class classifier instead, we retained the middle three classes as a check on the informational content of the important fragments.
 
 The performance of the classifiers is not exceptional on average but is certainly better than random, which provides further validation for our hypothesis that information is contained simply in the fragment composition of the molecules. The average area under the curve (AUC) for prediction of occupancy in class 0 (non-permeating) is $0.853\pm0.003$, while the AUC for prediction of occupancy in class 4 is $0.744\pm0.005$, significantly above the AUC of a random classifier, which is 0.5.  We note that predictivity is overall higher for class 0, probably due to imbalance in the training data (almost 50\% of training examples in class 0, only $\approx$20\% in class 4). The enrichment of class 4 at a percentage level $\rho_{0.1} = 100*0.1/N = 10\%$ is on average \rood{$118\pm5 \%$}, indicating that the classifier is able to find twice as many hits as if it chose at random, while the maximum enrichment at $10\%$ of any of the classifiers is $275\%$, although a few classifiers on their own do perform no better than random (cf.\ \blauw{Fig.\ S3}). In addition, the on-average monotonic decrease of the enrichment with $m$ demonstrates that the probability rankings are sensible.  Overall, the enrichment indicates that by taking the top ten percent most probable molecules as predicted by any given classifier, one should be able to on average find twice as many hits as if one were to select compounds at random, which represents a significant shrinkage of the search space.  We also expect that the overall performance would be improved by model voting over multiple classifiers, as discussed earlier in the \blauw{Results} section.

\section{Conclusions}
\label{sec:conclusions}
Remarkably, by combining traditional machine learning approaches with a novel, fragment-based molecular description inspired by natural language n-grams--with no \emph{a priori} expert input--we are able to demonstrate a new algorithm that has the potential to identify the chemical submolecules that impart compounds with desirable properties and identify existing drugs with those properties, leading to a unique avenue for hybrid drug design and drug reuse. Specifically, we have employed our algorithm to identify a set of \rood{forty-three} fragments expected to confer on drugs the ability to permeate the outer membrane of \emph{P.\ aeruginosa}, as well as \rood{eight} compounds expected to be good outer membrane permeators, of which thus far \rood{five} have been directly experimentally validated.  This work represents an important step forward for rational hybrid drug design, particularly for antibiotics against Gram-negative bacteria.

\section{Acknowledgements}
This work was supported by NIAID/NIH grant number R01AI136799.  RAM acknowledges a Los Alamos Director's Postdoctoral Fellowship. We thank Olga Lomovskaya, Qpex Biopharma for providing the Rempex compounds. We thank Paolo Ruggerone and Giuliano Malloci for valuable discussions.  We thank Illia S.\ Affanasiev for technical assistance with the MIC measurements. We thank Dr.\ Keith Haynes and Dr.\ Napoleon D'Cunha for synthesis of certain compounds used in the analysis. 

\section{Author Contributions}

R.A.M., N.W.H., V.V.R., H.I.Z., and S.G.\  conceived of the presented idea. R.A.M.\ developed the algorithm, wrote code, and performed the computational experiments. I.V.L.\ and J.M.\ performed experiments and analyzed MICs for different subsets of compounds and contributed equally. N.W.H.\ provided feedback on statistical tests. J.W.\ provided part of the chemical library for analysis. C.A.L.\, H.I.Z., and J.W.\ provided biochemical feedback on the identified fragments and hits. S.G.\ and H.I.Z. supervised the project. R.A.M., C.A.L., H.I.Z., and S.G.\  wrote the manuscript with feedback from all other authors.

\section{Supporting Information Available}

We provide the table of predicted hits along with their measured MIC values (if available) and the fraction of times they were reported out of all runs in the file \texttt{anc/projected\_hits.xlsx}.  We provide the table of active fragments along with the fraction of times they were reported out of all runs in the file \texttt{anc/topfragments.xlsx}. We also include a discussion of the theory behind multinomial logistic regression and three supplemental figures.

\section{Data Availability}

The code and example scripts for the Hunting Fox algorithm, the input train/test data, and the external library of 30,929 molecules used to identify hits not included in the train/test data are available upon request.

\bibliographystyle{unsrt}
\bibliography{main.bbl}

\end{document}


\maketitle

\renewcommand\thefigure{S\arabic{figure}}    
\setcounter{figure}{0}   
\renewcommand\thesection{S\arabic{section}}

\begin{table}[]
\centering
\caption{List of unique fragments of radius 1 and unique fragments of radius 2 contained in OU-31237.}
\begin{tabular}{ll}
\toprule
\textbf{SMILES} & \textbf{Radius} \\
\midrule
C=O & 1 \\
CC(N)=O & 1 \\
CCc & 1 \\
CF & 1 \\
CNC & 1 \\
CO & 1 \\
cCN & 1 \\
cOC & 1\\
cc(c)C & 1 \\
cc(c)O & 1 \\ 
ccc & 1 \\
cC(F)(F)F & 1 \\
cC(F)(F)F & 2 \\ 
cc(c)C(F)(F)F & 2 \\ 
ccc(cc)C(F)(F)F & 2 \\
cccc(c)C & 2 \\
ccccc & 2 \\
ccc(cc)CN & 2 \\
cc(C)cc(c)C & 2 \\
cc(c)CNC & 2 \\
cCNC(C)=O & 2 \\
cCC(=O)NC & 2 \\
CC(N)=O & 2 \\
cc(c)CC(N)=O & 2 \\
ccc(cc)CC & 2 \\
cccc(c)O & 2 \\
ccc(cc)OC & 2 \\
cc(c)OC & 2 \\
COc & 2 \\
\bottomrule
\end{tabular}
\label{tab:6Cslibrary}
\end{table}

\begin{figure}
    \centering
    \includegraphics[width=\textwidth]{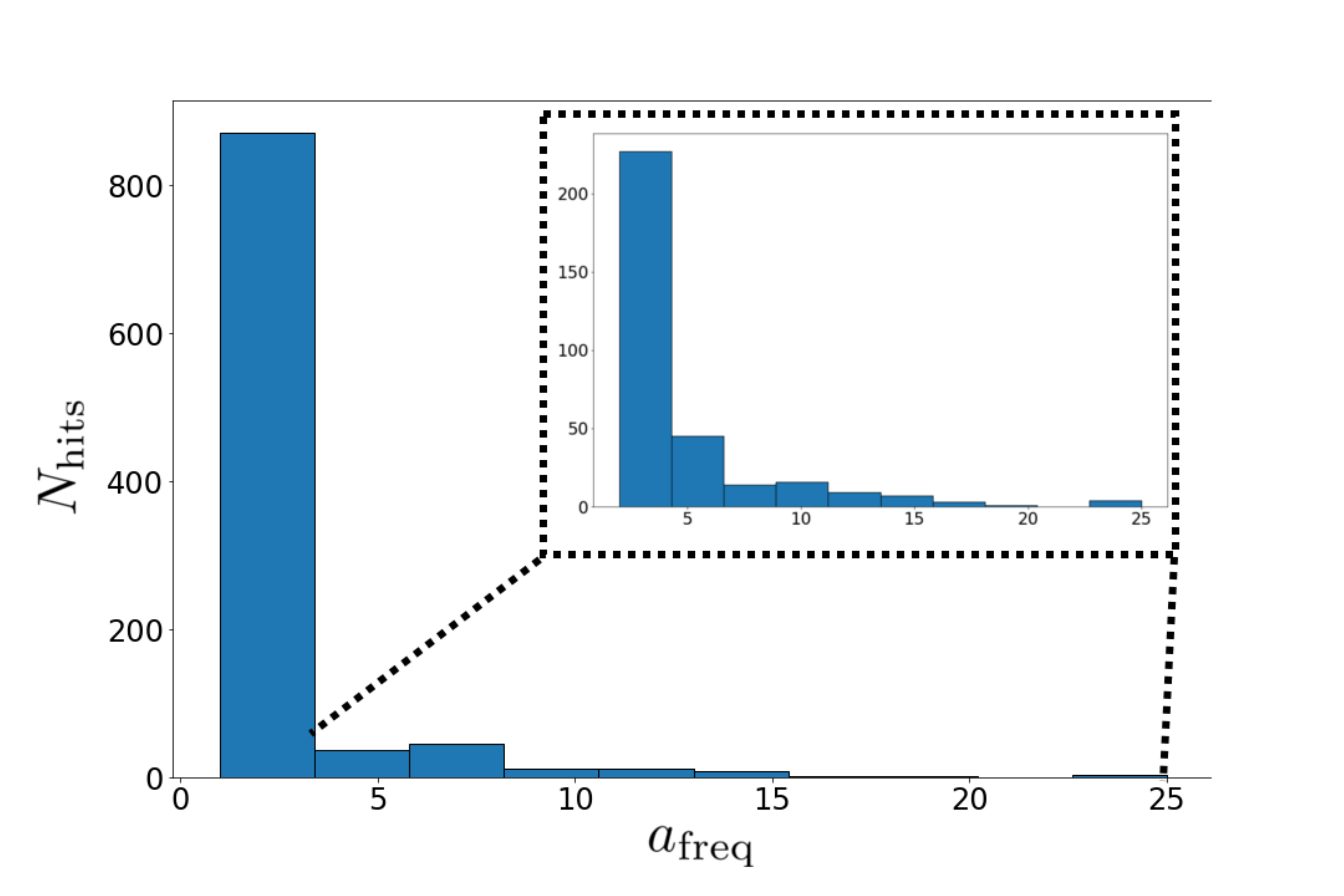}
    \caption{Histogram showing the number of compounds $N_\textrm{hits}$ identified by $a_\textrm{freq}$/\rood{ten} runs of Hunting FOX with randomly shuffled training/testing data. Inset shows the distribution for compounds that are identified by more than one run of the algorithm.}
    \label{fig:hitshists}
\end{figure}

\begin{figure}
    \centering
    \includegraphics[width=\textwidth]{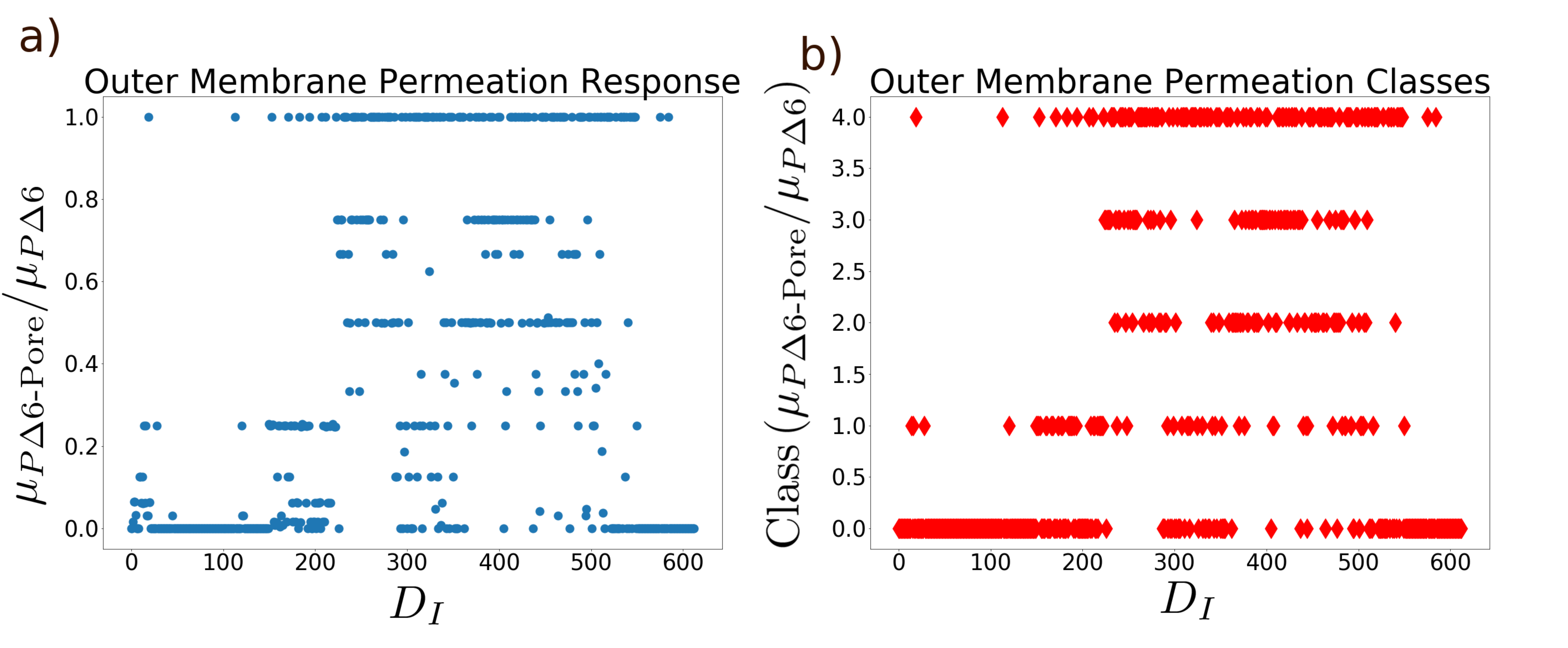}
    \caption{Illustration of class assignment for MIC ratio response variables. In a) we show the actual MIC ratio $\frac{\mu_{P\Delta6\textrm{Pore}}}{\mu_{P\Delta6}}$ for each compound considered in a training or testing simply versus its index $D_I$. In b) we show the corresponding class assignment. Note the discreteness of the data.}
    \label{fig:dataclasses}
\end{figure}

\section{Theory of multinomial logistic regression}
\label{subsec:mlreg}
We chose to fit our classifiers by employing the multinomial logistic regression technique, since it is well-known and therefore easily employed, provides an easily-interpretable model and, in addition to projected compound classes, the probabilities of a compound belonging to each class. It is a method employing maximum likelihood estimation to solve a multiclass problem in which, in essence, we seek to find the probability of a particular outcome being in class $C$, $C \in [0,4]$  given a molecular vector, $\vec{\ell}$ \cite{Hosmer2013AppliedRegression.},

\begin{equation}
    P\left(\textrm{Class}\left(\frac{\mu_{P\Delta 6\textrm{-Pore}}}{\mu_{P \Delta 6}}\right) = C|\vec{\ell}\right) =
    \begin{cases}
    1/\sum_{k=0}^4  \exp\left(g_k(\vec{\ell}), \right)  & C=0\\
     \exp\left(g_C(\vec{\ell}) \right)/\sum_{k=0}^4 \exp\left(g_k(\vec{\ell}), \right) & \textrm{else}
    \end{cases}
\end{equation}
where $g_C(\vec{\ell})$ is the logit function, which is the particular form of the fit chosen in general logistic regression, \cite{Hosmer2013AppliedRegression.}
\begin{equation}
    g_C(\vec{\ell}) = \ln\left[\frac{P(\textrm{Class}\left(\frac{\mu_{P\Delta 6\textrm{-Pore}}}{\mu_{P \Delta 6}}\right) = C|\vec{\ell})}{P(\textrm{Class}\left(\frac{\mu_{P\Delta 6\textrm{-Pore}}}{\mu_{P \Delta 6}}\right)=0|\vec{\ell})} \right] = \beta_{C0} + \sum_{k=1}^p\beta_{Ck}\ell_k,
\end{equation}
where $\beta_{Ck}$ are a set of coefficients to be fitted, and $m$ is the length of the molecular descriptor vector. The logit expresses the log of the ratio of the probabilities of being in class $C$ versus class 0 \cite{Hosmer2013AppliedRegression.}. Employing this expression of the probabilities results in the following expression for the log-likelihood function \cite{Balakumar2016GlmnetPython},
\begin{equation}
    \mathcal{L} = - \left[\frac{1}{N}\sum_{i=1}^N \left(\sum_{k=1}^K I(g_i = C) \left(\beta_{0k} + \ell_i^T \beta_k \right) - \ln\left(\sum_{k=1}^K \exp\left(\beta_{0k} + \ell_i^T \beta_k \right) \right)\right) \right],
\end{equation}
where $I(g_i = \mathcal{L})$ is an indicator function, and $\beta_k$ is the $k$th column of the $p \times K$ coefficient matrix, where $K$ is the number of classes and $p$ is the length of the molecular descriptor vector $\vec{\ell}$.  Then multinomial logistic regression is the procedure of maximizing the log-likelihood of this function, potentially subject to additional constraints such as regularization for sparsification.  Specifically, the \texttt{cvglmnet} function solves the optimization problem,
\begin{align}
\begin{split}
    \min_{\{\beta_{0k},\beta_k)\}_1^K} & - \left[\frac{1}{N}\sum_{i=1}^N \left(\sum_{k=1}^K I(g_i = C) \left(\beta_{0k} + \ell_i^T \beta_k \right) - \ln\left(\sum_{k=1}^K \exp\left(\beta_{0k} + \ell_i^T \beta_k \right) \right)\right) \right] \\
   & +\lambda \left[(1-\alpha)||\beta||^2_F +\alpha \sum_{j=1}^p ||\beta_j||_q \right],
\end{split}
\end{align}
where $\alpha=1,q=2$ was used for sparse classification and $\alpha=0$ for non-sparse classification.
\begin{figure}
    \centering
    \includegraphics[width=\textwidth]{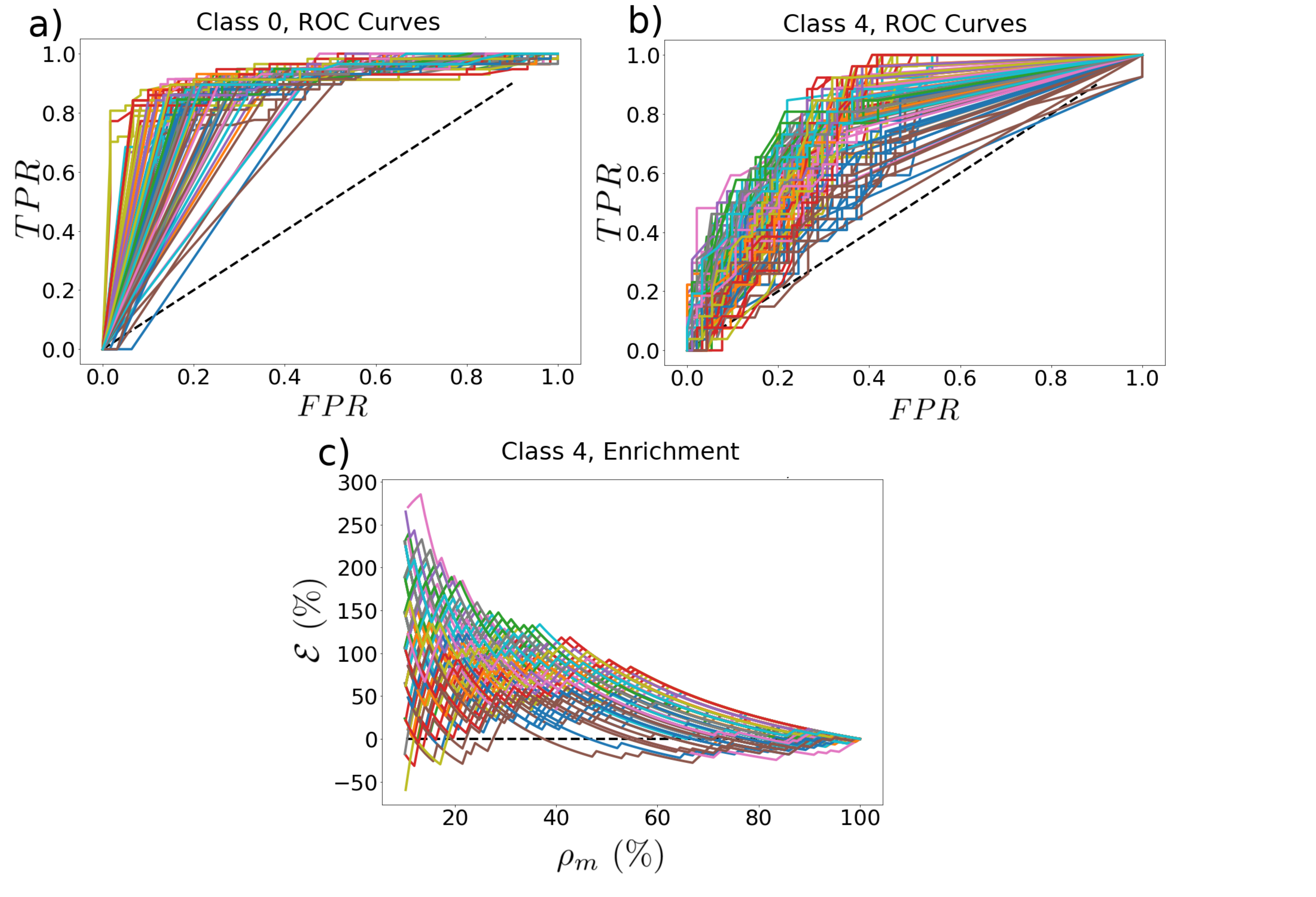}
    \caption{Individual ROC curves resulting from different classifiers for (a) class 0 and (b) class 4 and (c) individual enrichment ($\mathcal{E}$) versus percent  $\rho_m$ of molecules considered after sorting by predicted probabilites for class 4 occupancy. Different colors demonstrate curves resulting from different classifiers. Dashed black lines indicate the expected performance of a random classifier.}
    \label{fig:classperform}
\end{figure}
\bibliographystyle{unsrt}
\bibliography{bibliography.bib}